\documentclass[12pt]{article}
\usepackage{amsmath}
\usepackage{amssymb}
\usepackage{amsthm}
\usepackage{graphicx}
\usepackage{lineno}
\usepackage{float}
\usepackage[all]{xy}
\usepackage{mathrsfs}

\usepackage{xcolor}

\newtheorem{definition}{Definition}

\begin{document}

\title{\bf Neurons as hierarchies of quantum reference frames}

\author{{Chris Fields$^a$\footnote{Corresponding author at: 23 Rue des Lavandi\`{e}res, 11160 Caunes Minervois, FRANCE; {\it E-mail address}: fieldsres@gmail.com}, James F. Glazebrook$^b,^c$ and Michael Levin$^d$}\\ \\
{\it$^a$ 23 Rue des Lavandi\`{e}res, 11160 Caunes Minervois, FRANCE}\\
{ORCID: 0000-0002-4812-0744} \\
{\it$^b$ Department of Mathematics and Computer Science,} \\
{\it Eastern Illinois University, Charleston, IL 61920 USA} \\
{\it$^c$ Adjunct Faculty, Department of Mathematics,}\\
{\it University of Illinois at Urbana-Champaign, Urbana, IL 61801 USA}\\
{\it$^d$ Allen Discovery Center at Tufts University, Medford, MA 02155 USA}\\
{ORCID: 0000-0001-7292-8084}
}

\maketitle

{\bf Abstract} \\
Conceptual and mathematical models of neurons have lagged behind empirical understanding for decades.  Here we extend previous work in modeling biological systems with fully scale-independent quantum information-theoretic tools to develop a uniform, scalable representation of synapses, dendritic and axonal processes, neurons, and local networks of neurons.  In this representation, hierarchies of quantum reference frames act as hierarchical active-inference systems.  The resulting model enables specific predictions of correlations between synaptic activity, dendritic remodeling, and trophic reward.  We summarize how the model may be generalized to nonneural cells and tissues in developmental and regenerative contexts. \\


\section{Introduction}

Neurons are canonical biological information processors.  Theoretical and, particularly, conceptual models of neural information processing, however, lag increasingly far behind our developing empirical understanding of neurons as electrically-excitable cells.  Experimental work over the past two decades has, for example, firmly established that dendrites undergo activity-dependent remodeling \cite{butz:09, carulli:11, hogan:20}, particularly alterations of spine location, density, and function \cite{runge:20}, even in adult humans.  This ontogenic process is functionally analogous to the evolution of structural and positional diversity of dendrites as they have adapted to a spectrum of functional roles \cite{wittenberg:16}, e.g. the implementation of deep learning via synaptic plasticity \cite{guerguiev:17,sardi:18}.   Neurons are not, therefore, static structures, but rather can be regarded as undergoing continuous development throughout the life cycle.  This dynamic process has significant consequences for both neuron-level and organism-level function.  For example, in organisms (such as caterpillars transitioning into butterflies or moths) that exhibit drastic remodeling and reconstruction of their brains, some of their learned memories remain and survive the process \cite{blackiston:15}.  In other contexts, memories can be imprinted on newly regenerating brains from other tissues \cite{shomrat:13, mcconnell:67}, underscoring the plasticity of large-scale neural structure and of the information stored within it. These effects of remodeling are not just an issue for so-called lower animals, as applications in regenerative medicine are likely to soon produce human patients in whom some portion of the brain has been replaced by the progeny of na\"{i}ve stem cells to treat degenerative disease or brain damage.

It has also been shown that subnetworks of dendrites can compute local logical operations including exclusive-or (XOR) \cite{gidon:20}.  Nonetheless, the idea that dendritic trees effectively compute weighted sums, dating back to \cite{mp:43}, continues not only to dominate artificial neural network (ANN) development, but also to guide biological thinking.  Explicit cable-theory modeling can reproduce time-dependent signal propagation and processing in simplified representations of dendritic trees \cite{segev:00}, but rapidly becomes intractable with increasing geometric complexity.   Increasingly sophisticated hardware models of neurons allow exploration of phase coding, frequency modulation, and other forms of hybrid analog-digital computing \cite{schuman:17, tang:19}, but are not readily mapped to biological neural networks and are not standard components of a theoretical neuroscientist's toolkit.

Conceptualizing neurons as biological wires -- even functionally sophisticated wires -- provides, moreover, no insight into the fundamental question of cognitive neuroscience: the question of how information present in the external world is encoded, via some combination of evolution, development, remodeling, and learning, into the functional architecture of a nervous system.  Brains are not constructed, as ANNs are, in the absense of functional activity, nor do they ``begin learning'' from a default initial state of uniform connectivity.  Brains are not ``wired up'' into some starting configuration, after which they are turned on and exposed to the world.  Brains instead develop from neural primordia that are already functionally complex multicellular microenvironments.  The function of the nervous system, from its earliest phylogenetic and ontogenetic role in directing morphogenesis \cite{fbl:20} to adult cognition, depends on the ability of individual neurons to negotiate this microenvironment, both structurally and functionally.  The local microenvironment, with its diverse cell types, biochemical and (micro)anatomical complexity, and network of biomolecular and bioelectric signaling, is the ``world'' with which each individual neuron interacts, and is each neuron's sole source of nongenomic information.  There are many interesting scenarios supporting this viewpoint; some are of the `socialization' and decision making type.  For instance, assortative neural networks demonstrate collective resilience, whereby nodes of a certain degree have an affinity to team up with those of the same topological type, and thus often contribute to formation of a small world network structure \cite{barrat:08,sporns:10}. Neural network modeling of independent tasks in the prefrontal cortex in \cite{latham:05}, for example, reveals how seemingly separate neuronal groups engaged in their respective tasks may sometimes be drawn into a coalition, casting votes for a stimulus, and then enacting a committee-like decision.

In line with the Free Energy Principle (FEP) and the idea of Bayesian active inference \cite{friston:10a,friston:13a,friston:07a,ramstead:20,friston:19}, individual neurons can be considered cognitive agents that minimize aggregate uncertainty about their future states (variational free energy or VFE) by actively exploring and developing predictive models of their local microenvironments.  It is the joint activity of hundreds (in {\em C. elegans} \cite{varshney:11}) to tens of billions (in primates \cite{herculano:11}) of neurons within this jointly-constructed environment that encodes information sourced from the external world.  Hence an adequate conceptual model of neurons as biological systems must explain how each neuron's construction of a predictive model of its local microenvironment results in the joint encoding of a predictive model of the external environment at the scale of the entire brain.  The utility of FEP based models of cell sorting in morphogenesis \cite{friston:15, kuchling:20}, cortical minicolumn and local-circuit organization \cite{kiebel:11, bastos:12, shipp:13, kanai:15, adams:15}, and network and whole-brain function \cite{friston:13a, clark:13, hohwy:13, seth:13, friston:15a} suggests that a scale-free conceptual framework is the right way to approach this question of functional coherence between microenvironment models constructed by individual neurons and external environment models constructed by brains.  The FEP is, however, fundamentally a statement about error minimization; it is an implementation-independent principle of thermodynamics \cite{friston:19}.  It does not tell us anything specific about neuronal functional architecture, or about {\em how} neurons interact with either each other or the non-neuronal components of their microenvironments.  

Here we suggest that the idea of a {\em quantum reference frame} (QRF) developed within quantum information theory \cite{aharonov:84, bartlett:07} is usefully applied to characterize neurons both architecturally and functionally.  While quantum information theory employs the formalism of standard quantum theory, it makes no scale-dependent assumptions and is applicable from sub-microscopic to cosmological scales.  The formalism of QRFs, in particular, is applicable at any scale.  A QRF is a physical system that assigns units of measurement, and hence an operational semantics, to each observational outcome, enabling outcomes obtained at different times or places, or with different QRFs, to be compared.  Macroscopic systems such as meter sticks, clocks, gyroscopes, and the Earth with its gravitational and magnetic fields are canonical examples of QRFs.  While reference frames are typically thought of as fully-specifiable abstractions in classical physics, this is not the case in quantum theory: QRFs are physical systems that encode unmeasurable but functionally relevant quantum phase information.  Every QRF is, therefore, in an important sense unique.  A QRF cannot, in particular, be fully specified either structurally or functionally by any finite bit string; it is ``nonfungible'' in the terminology of \cite{bartlett:07}.  Alice can, in this case, share a QRF with a distant observer Bob only by physically transferring the QRF to Bob; sending any finite description is provably insufficient.  Alice can, moreover, transfer her QRF to Bob successfully only if Bob already possesses a functionally equivalent QRF \cite{fm:19}; Alice cannot, for example, successfully communicate the meaning of ``1 meter'' to Bob unless Bob already has a QRF that effectively functions as a meter stick.  From an information processing point of view, a QRF is a quantum computer: it outputs an outcome value that is reproducible within some fixed resolution, and that is assigned standardized units that confer an operational semantics, when given a ``raw'' measurement, or simply a physical interaction, as input.

We propose, in particular, that neurons are usefully regarded as hierarchies of QRFs.  Neurons in this representation are hierarchical measurement devices that sample their surrounding cellular microenvironments at multiple scales and encode scale-specific expectations about how those microenvironments will behave.  The output of a particular neuron encodes, for each cell that receives it, a specific, nonfungible representation of the transmitting neuron's local measurements.  These representations are encoded in the ``units of measurement'' defined by the sending neuron's particular combination of QRFs.  They impose operational semantic constraints on all downstream processing.  These constraints are inherited by, and give a particular operational meaning to, the activity patterns of the neural system as a whole.

We develop this proposal using a general formalism, that of Barwise-Seligman classifiers \cite{barwise:97} organized into networks with well-defined limits and colimits (cone-cocone diagrams or CCCDs) developed in \cite{fg:19a} and applied to human cognitive architecture in \cite{fg:19b, fg:20, fg:21}.  Barwise-Seligman classifiers are naturally interpreted as representing either logical or probabilistic constraints; hence CCCDs represent maximal collections of mutually-consistent constraints, and indeed generalize hierarchical Bayesian inference \cite{fg:21}.  To model neurons as hierarchies of QRFs is to represent them as 1) quantum computers that 2) implement hierarchical Bayesian inferences determined by 3) their specific three-dimensional (3d) geometries and 4) the differential signal-transduction velocities that their geometries impose.  This formalism extends naturally to networks of neurons, at any scale from a local circuit -- e.g. a minicolumn -- up to an entire brain.

We begin in \S\ref{QRFs} by briefly reviewing the implementation of QRFs by generic quantum systems \cite{fm:20, fg:20a, fgm:21} and their representations by CCCDs.  We then turn to biological systems, reviewing the representation of simple homeostatic setpoints as QRFs \cite{fl:20, fgl:21} and extending this treatment to perisynaptic processes in neurons in \S\ref{macromolecular}.  This allows us in \S\ref{dendrites} to represent the hierarchical structures of dendrites as hierarchical QRFs that detect the states -- activity patterns -- of particular ``objects'' -- spatially-organized aggregates of presynaptic axon terminals -- in their microenvironments.  We then extend this representation to neurons in \S\ref{neurons}, showing how the integration of signals from multiple dendritic branches implements an effectively tomographic computation.  This view of neurons as state identifiers scales upwards to functional networks, the global activity of which provide coarse-grained representations of particular components of the external environment.  We then briefly review in \S\ref{cells} results from a variety of systems showing that with suitable spatial and temporal scaling, this model applies to electrically-excitable cells generally.  We consider simulation modeling and experimental approaches suggested by this framework in \S\ref{conclusion}.

\section{QRFs and their representation by CCCDs} \label{QRFs}

\subsection{QRFs as generic representations of measurement}

Information processing in biological systems has traditionally been represented as classical; despite theoretical arguments from a variety of perspectives \cite{schro:44, HP:96, bord:13, tononi:15, georgiev:20} and experimental evidence for the functional relevance of quantum coherence in photoreception and magnetoreception \cite{arndt:09, lambert:12, melkikh:15}, quantum biology remains in its infancy \cite{marais:18, cao:20,brookes:17,mcfadden:18}.  Motivated in part by recent analyses showing that cellular bioenergetic resources fall orders of magnitude short of those needed for fully-classical computation at macromolecular scales \cite{fl:21}, here we adopt a quantum-theoretic perspective from the start.  This perspective allows full use of the quantum information toolkit, including the concept of a QRF; it offers the advantages of full generality and applicability at any scale \cite{fgl:21}.

\begin{figure}[H]
\centering
\includegraphics[width=13 cm]{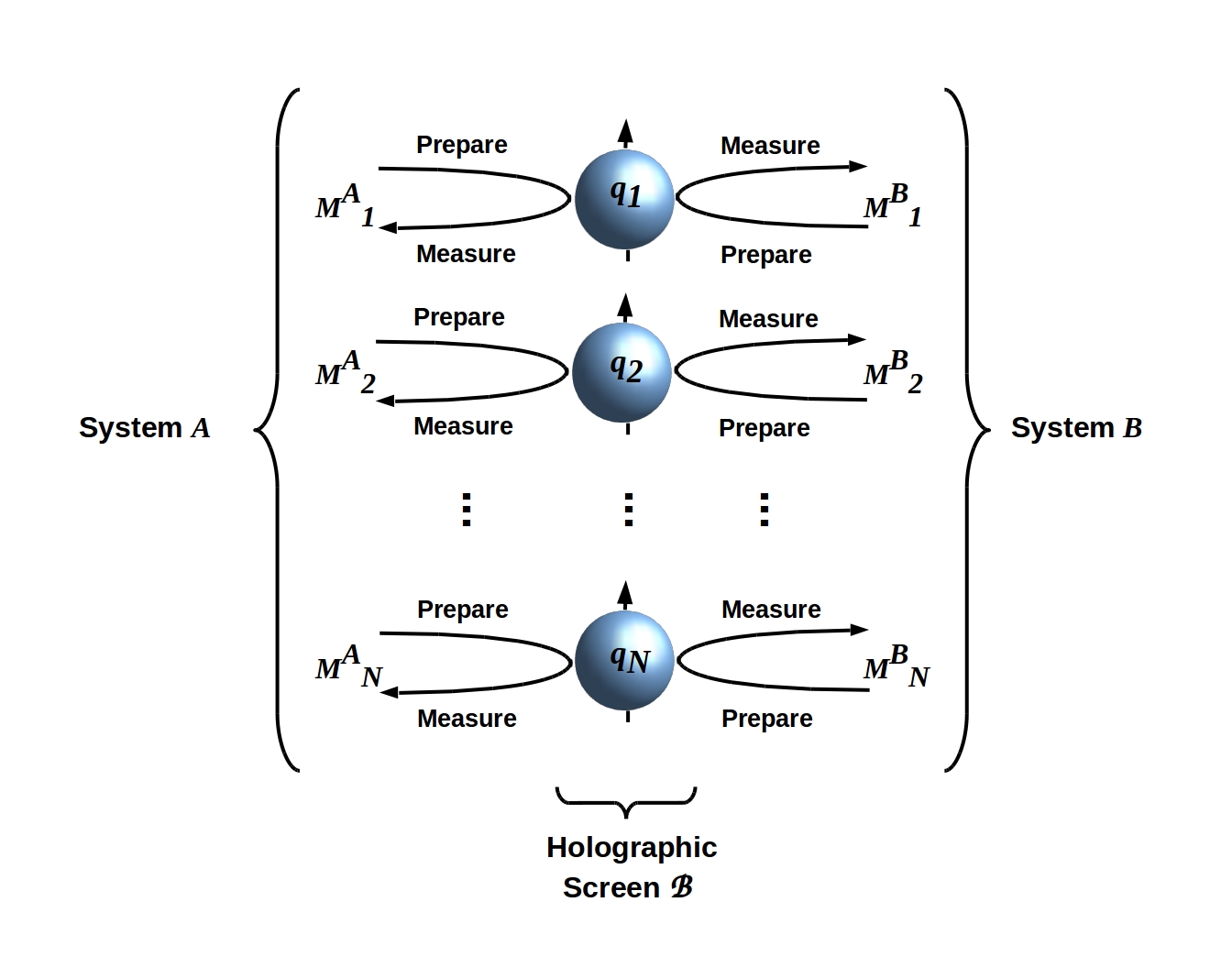}
\caption{A holographic screen $\mathscr{B}$ separating systems $A$ and $B$ with an interaction $H_{AB}$ given by Eq. \eqref{ham} can be realized by an ancillary array of noninteracting qubits that are alternately prepared by $A$ ($B$) and then measured by $B$ ($A$).  There is no requirement that $A$ and $B$ share preparation and measurement bases, i.e. QRFs.  Adapted from \cite{fg:21} Fig. 1, CC-BY license.}
\label{qubit-screen-fig}
\end{figure}

Consider an isolated, finite, bipartite system $AB$ for which the interaction $H_{AB}$ is weak enough and the time interval of interest short enough that considering $AB$ to be separable, i.e. having a joint state $|AB \rangle$ that factors as $|AB \rangle = |A \rangle |B \rangle$, is a good approximation.  In this case, it is always possible to choose a basis in which the interaction can be written:

\begin{equation} \label{ham}
H_{AB} = \beta^k k_B T^k \sum_i^N \alpha^k_i M^k_i,
\end{equation}
where $k = ~A$ or $B$, the $M^k_i$ are $N$ Hermitian operators with eigenvalues in $\{ -1,1 \}$, the $\alpha^k_i \in [0,1]$ are such that $\sum^N_i \alpha^k_i = 1$, and $\beta^k \geq$ ln 2 is an inverse measure of $k$'s thermodynamic efficiency that depends on the internal dynamics $H_k$ \cite{fm:19, fm:20, fl:20}.  In the form given by Eq. \eqref{ham}, the interaction can, without loss of generality, be regarded as defined at a holographic screen $\mathscr{B}$ separating $A$ from $B$; the operators $M^k_i$ can in this case be regarded as ``preparation'' and ``measurement'' operators that alternately write and read bit values encoded on $\mathscr{B}$ \cite{fm:20, addazi:21}.  The screen $\mathscr{B}$ can be realized as an ancillary qubit array as in Fig. \ref{qubit-screen-fig}.  The thermodynamic factor $\beta^k k_B T^k$ in Eq. \eqref{ham} assures compliance with {\em Landauer's Principle} \cite{landauer:61, landauer:99, bennett:82}, i.e. assures that the per-bit free-energy cost of classical bit erasure is paid on each cycle (see \cite{fg:20, fgl:21} for discussion).

Now consider $A$ to be an ``observer'' that decomposes $B$ into a ``system of interest'' $S$ and its surrounding ``environment'' $E$.  As a familiar example, $S$ may be some item of laboratory apparatus and $E$ the surrounding laboratory.  We suppose further that $A$ is primarily interested in the state $|P \rangle$ of some proper component $P$ of $S$, conventionally called the ``pointer'' of $S$, that is separable from the remaining proper component $R$ (the ``reference'' component), i.e. $S = PR$ and $|PR \rangle = |P \rangle |R \rangle$ over the observation times of interest.  Under these conditions, repeated observations of a fixed state $|R \rangle$ of $R$ allow the {\it identification} of $S$ as a system; in our example, repeated observations of the fixed size, shape, location, etc. of an item of laboratory apparatus allow its identification over time and hence unambiguous observations of its pointer state \cite{fg:20a, fgm:21}.  These repeated observations effect decoherence of $S$, and hence enforce separability of $S$ from $E$ \cite{fm:20}.  Nothing in the above changes when pure states $|P \rangle$ and $|R \rangle$ are replaced by densities $\rho_P$ and $\rho_R$, interpreted as distributions over time series of pure states, with the separability condition $\rho_{PR} = \rho_P \rho_R$.

Assigning mutually-exclusive subsets of bits encoded on $\mathscr{B}$ to $P$, $R$, and $E$ is clearly a computational process that must be implemented by the internal interaction $H_A$ of $A$.  Indeed this process is completely independent of $H_B$ or, equivalently, of the decompositional structure of $B$ \cite{fg:20a}.  Tracking the states of $P$, $R$, and $E$ over time requires an adequate memory resource and a comparison function capable of detecting changes at some suitably coarse-grained resolution.  As shown in \cite{fg:20a, fgm:21}, these computations can be regarded as implementing QRFs for $P$, $R$, and $E$, the functions of which can, without loss of generality, be specified by networks (formally, cocones) of Barwise-Seligman classifiers as described in \S\ref{classifiers} below.

It is worth emphasizing here that the decomposition $B = PRE$ is not ``objective'' in any sense; the components $P$, $R$, and $E$ are defined, relative to and hence ``for'' $A$, by their respective QRFs.  These QRFs can, therefore, themselves be labeled `$P$', `$R$', and `$E$' without ambiguity.  Similarly, the states $|P \rangle$, $|R \rangle$, and $|E \rangle$ encoded on $\mathscr{B}$ in the basis specified by Eq. \eqref{ham} are not ``objective'' but rather defined as the domains, respectively, of the QRFs $P$, $R$, and $E$.  They are, therefore, also strictly relative to $A$ (cf. the characterization of quantum states as {\it personal} in \cite{fuchs:13, mermin:19} or as {\em observer-relative} in \cite{rovelli:96}).  The computations implemented by $P$, $R$, and $E$ result in classical, i.e. irreversible encodings of state information to $A$'s memory; therefore they require free energy as discussed in \cite{fg:20a}.  This free energy can only come from $B$; hence some fraction of the bits encoded on $\mathscr{B}$ must be viewed as supplying the free energy required for computation.  The values of these bits cannot, therefore, affect the computational outcomes of     $P$, $R$, and $E$; they are effectively traced over.  This obligate trace operation renders the outputs of $P$, $R$, and $E$ coarse-grained representations $\rho_P$, $\rho_R$, and $\rho_E$ that can be viewed as probability distributions, and effectively as weighted averages, in the basis specified by Eq. \eqref{ham}, or as pure states in a ``computational basis'' with reduced dimension.  The dimension reduction or coarse-graining induced by the action of any QRF, solely in consequence of its free-energy requirements, enables the construction of QRF hierarchies with distinguishable layers as discussed in the case of neurons in \S\ref{dendrites} and \S\ref{neurons} below.

\subsection{Commutativity constraints on QRFs} \label{commutativity}

A system $S$ is only useful, in practice, as a measurement apparatus if it can be identified over macroscopic time, i.e. has a stable coarse-grained reference state $\rho_R$.  ``Stability'' requires, in particular, that observing the environment $E$ cannot affect $\rho_R$; hence the QRFs $R$ and $E$ must commute.  The reference state $\rho_R$ that allows identification of $S$ must, moreover, remain stable when the pointer component $P$ is observed; hence $R$ and $P$ must commute.  Similarly $P$ must commute with $E$.  These commutativity conditions are, effectively, consequences of separability: they follow directly from the assumption that $P$, $R$, and $E$ are each distinguishable from the others.

These commutativity conditions on the QRFs $P$, $R$, and $E$ can be summarized in diagrammatic form by:

\begin{equation}\label{diag1}
\begin{gathered}
\xymatrix{
\rho_P \ar@{-->}[r]^{\mathbb{M}_P} & \rho^{\prime}_P & \rho_R \ar@{-->}[r]^{\mathbb{M}_R} & \rho^{\prime}_R & \rho_E \ar@{-->}[r]^{\mathbb{M}_E} & \rho^{\prime}_E \\
|\mathscr{B} \rangle \ar[u]^{P}  \ar[r]^{\mathcal{P}_U} & |\mathscr{B} \rangle^{\prime} \ar[u]_{P} & |\mathscr{B} \rangle \ar[u]^{R}  \ar[r]^{\mathcal{P}_U} & |\mathscr{B} \rangle^{\prime} \ar[u]_{R} & |\mathscr{B} \rangle \ar[u]^{E}  \ar[r]^{\mathcal{P}_U} & |\mathscr{B} \rangle^{\prime} \ar[u]_{E}
}
\end{gathered}
\end{equation}
~ \\
\noindent
where $|\mathscr{B} \rangle$ is the bit string encoded on $\mathscr{B}$ by the action of $H_{AB}$, $\mathcal{P}_U$ is the time propagator of the joint system $U = AB$, and $\mathbb{M}_P$, $\mathbb{M}_R$, and $\mathbb{M}_E$ are Markov kernels on the state spaces of $\rho_P$, $\rho_R$, and $\rho_E$, respectively.  The boundary $\mathscr{B}$ functions as a {\em Markov Blanket} (MB) \cite{pearl:88, clark:17} that renders the system state $|A \rangle$ {\em conditionally independent of} $|B \rangle$ at the ``microscale'' defined by $H_{AB}$ \cite{fm:20b}, while $\rho_P$, $\rho_R$, and $\rho_E$ are effective MB states at the coarse-grained ``computational scale'' at which $A$ writes observed state information to memory.  The probabilities encoded by the $\mathbb{M}_P$, $\mathbb{M}_R$, and $\mathbb{M}_E$ are posterior probabilities for $A$ from a Bayesian perspective; the task of $A$ as an observer implementing active inference \cite{friston:10a,friston:13a,friston:07a,ramstead:20,friston:19} is to learn priors, or equivalently, to implement actions on $B$, that predict $\mathbb{M}_P$, $\mathbb{M}_R$, and $\mathbb{M}_E$ as accurately as possible.

The operators $\mathbb{M}_P$, $\mathbb{M}_R$, and $\mathbb{M}_E$ are, clearly, Markovian if but only if the joint density $\rho_{PRE}$ factors as $\rho_{PRE} = \rho_P \rho_R \rho_E$, i.e. if but only if $P$, $R$, and $E$ can be assumed to be separable.  As noted above, separability can be assumed only in the limit of weak interactions and/or short observation times.  Hence Markovian evolution of $P$, $R$, and $E$ is an assumption valid in this limiting case, not a fact.  Evolution being Markovian corresponds, by definition, to probabilities satisfying the Kolmogorov axioms and hence to the Bell \cite{bell:64} and Leggett-Garg \cite{emary:14} inequalities being satisfied, not violated.  Hence physically, the Markovian evolution corresponds to phase correlations being negligible, i.e. the absence of quantum entanglement or changes of ``intrinsic'' measurement context \cite{kochen:67, mermin:93, dzh:18} as discussed further below.  These conditions can be summarized as observations being ``sufficiently coarse-grained'' at the computational scale to be treated as classical.

The commutativity conditions given by Eq. \eqref{diag1} are instances of the general condition required to interpret any physical process as computation \cite{horsman:14}.  We can, therefore, regard $\rho_P$, $\rho_R$, and $\rho_E$ as encoding a ``semantic'' representation of the microscale states $\vert P \rangle$, $\vert R \rangle$, and $\vert E \rangle$ encoded on $\mathscr{B}$.  It is with respect to this completely-general notion of semantics that hierarchies of QRFs can be considered semantic, and hence virtual machine (VM) hierarchies \cite{smith:05}, in \S\ref{dendrites} - \ref{neurons} below.

\subsection{QRFs as constraint networks} \label{classifiers}

In recent work \cite{fg:19a,fg:19b,fg:20,fg:21} we have drawn extensively upon the semantically enriched Channel Theory of information flow developed in \cite{barwise:97}, outlining many examples and applications (see also in particular \cite{fg:20a,fgm:21}). Here we summarize the basic ideas, starting with that of a (Barwise-Seligman) ``classifier'' (or ``classification''), which categorically accommodates a ``context'' in terms of its constituent ``tokens'' in some language and the ``types'' to which they belong.

\begin{definition}\label{class-def-1}
A {classifier} $\mathcal{A}$ is a triple $\langle Tok(\mathcal{A}), Typ(\mathcal{A}), \models_{\mathcal{A}} \rangle$ where $Tok(\mathcal{A})$ is a set of ``tokens'', $Typ(\mathcal{A})$ is a set of ``types'', and $\models_{\mathcal{A}}$ is a ``classification'' relation between tokens and~types.
\end{definition}
\noindent
Note that this definition specifies a classifier/classification as an object in the category of Chu spaces \cite{barr:79,pratt:99a,pratt:99b} where `$\models_{\mathcal{A}}$' is realizable by a satisfaction relation valued in some set $\mathbf{K}$ having no structure assumed.

Morphisms between classifiers are specified by the following:

\begin{definition}\label{class-def-2}
Given two classifiers $\mathcal{A} = \langle Tok(\mathcal{A}), Typ(\mathcal{A}), \models_{\mathcal{A}} \rangle$ and $\mathscr{B} = \langle Tok(\mathscr{B}), Typ(\mathscr{B}), \models_{\mathscr{B}} \rangle$, an {infomorphism} $f: \mathcal{A} \rightarrow \mathscr{B}$ is a pair of maps $\overrightarrow{f}: Tok(\mathscr{B}) \rightarrow Tok(\mathcal{A})$ and $\overleftarrow{f}: Typ(\mathcal{A}) \rightarrow Typ(\mathscr{B})$ such that $\forall b \in Tok(\mathscr{B})$ and $\forall a \in Typ(\mathcal{A})$, $\overrightarrow{f}(b) \models_{\mathcal{A}} a$ if and only if $b \models_{\mathscr{B}} \overleftarrow{f}(a)$.
\end{definition}
\noindent
This last definition can be represented schematically as the requirement that the following diagram commutes:

\begin{equation}\label{info-diagram-1}
\begin{gathered}
\xymatrix@!C=3pc{\rm{Typ}(\mathcal{A}) \ar[r]^{\overrightarrow{f}}   & \rm{Typ}(\mathscr{B}) \ar@{-}[d]^{\models_{\mathscr{B}}} \\
\rm{Tok}(\mathcal{A}) \ar@{-}[u]^{\models_{\mathcal{A}}}  & \rm{Tok}(\mathscr{B}) \ar[l]_{\overleftarrow{f}}}
\end{gathered}
\end{equation}
\noindent
Following these definitions, Channel Theory can be represented as a category comprising classifiers as objects and infomorphisms as arrows, which can be seen to be equivalent to the category comprising Chu spaces as objects and Chu morphisms as arrows \cite{barwise:97}.

Much of the formulism of \cite{barwise:97} revolves around the idea of a {\em distributed system} in which the classifiers and infomorphisms between them function as `logical gates' as further exemplified in \cite{fg:19a,fg:19b,fg:20,fg:21}. Significantly instrumental in this schemata is a finite, commuting {\em cocone diagram} (CCD) depicting a flow of infomorphisms sending inputs to a {\em core} $\mathbf{C^\prime}$ that is the colimit of the underlying classifiers if such exists:

\begin{equation}\label{ccd-1}
\begin{gathered}
\xymatrix@C=4pc{&\mathbf{C^\prime} &  \\
\mathcal{A}_1 \ar[ur]^{f_1} \ar[r]_{g_{12}} & \mathcal{A}_2 \ar[u]_{f_2} \ar[r]_{g_{23}} & \ldots ~\mathcal{A}_k \ar[ul]_{f_k}
}
\end{gathered}
\end{equation}
\noindent
There is a dual construction to this CCD, namely a commuting finite \emph{cone diagram} (CD) of infomorphisms on the same classifiers, where all arrows are reversed.  In this case the core of the (dual) channel is the limit of all possible downward-going structure-preserving maps to the classifiers $\mathcal{A}_i$.
Hence we can define the central idea of a finite, commuting {\em cone-cocone diagram} (CCCD) as consisting of both a cone and a cocone on a single finite set of classifiers $\mathcal{A}_i$ linked by infomorphisms as depicted below:

\begin{equation}\label{cccd-1}
\begin{gathered}
\xymatrix@C=6pc{&\mathbf{C^\prime} &  \\
\mathcal{A}_1 \ar[ur]^{f_1} \ar[r]_{g_{12}}^{g_{21}} & \ar[l] \mathcal{A}_2 \ar[u]_{f_2} \ar[r]_{g_{23}}^{g_{32}} & \ar[l] \ldots ~\mathcal{A}_k \ar[ul]_{f_k} \\
&\mathbf{D^\prime} \ar[ul]^{h_1} \ar[u]^{h_2} \ar[ur]_{h_k}&
}
\end{gathered}
\end{equation}
\noindent
Significantly, the CD functions as a ``memory-write system'' in Eq. \eqref{ccd-1}; this is used to represent the time-stamped ``write'' operations of a QRF in \cite{fg:20a,fgm:21}.  These diagrams can be extended to obtain a ``bow tie'' diagram as below in \eqref{cccd-3} which represents a coarse-graining of the semantics of $\mathcal{A}_i$ via $\mathbf{C^\prime}$, into a compressed representation $\mathcal{A}^{\prime}_i$.
\begin{equation}\label{cccd-3}
\begin{gathered}
\xymatrix@C=6pc{\mathcal{A}^{\prime}_1 \ar[r]_{g^{\prime}_{12}}^{g^{\prime}_{21}} & \ar[l] \mathcal{A}^{\prime}_2 \ar[r]_{g^{\prime}_{23}}^{g^{\prime}_{32}} & \ar[l] \ldots ~\mathcal{A}^{\prime}_j \\
&\mathbf{C^\prime} \ar[ul]^{h^{\prime}_1} \ar[u]^{h^{\prime}_2} \ar[ur]_{h^{\prime}_j}& \\
\mathcal{A}_1 \ar[ur]^{f_1} \ar[r]_{g_{12}}^{g_{21}} & \ar[l] \mathcal{A}_2 \ar[u]_{f_2} \ar[r]_{g_{23}}^{g_{32}} & \ar[l] \ldots ~\mathcal{A}_k \ar[ul]_{f_k}
}
\end{gathered}
\end{equation}
Diagrams such as \eqref{ccd-1} \eqref{cccd-3} can be further extended into hierarchical networks by adding intermediate layers of classifiers and infomorphisms. In this way they resemble artificial neural networks (ANNs), and in the ``bowtie'' form of Diagram \eqref{cccd-3}, they resemble variational autoencoders (VAEs) \cite{fg:19a}.  The core $\mathbf{C^\prime}$ in \eqref{cccd-3} can be viewed as both an ``answer'' computed by the $f_i$ from inputs to the $\mathcal{A}_i$ and, dually, as an ``instruction'' propagated by the $h_i$ (or in Diagram \eqref{cccd-3}, the $h^{\prime}_i$), to drive outputs from the $\mathcal{A}_i$ (or in Diagram \eqref{cccd-3}, the $\mathcal{A}^{\prime}_i$). This kind of dual input/output is precisely that of a QRF. Crucially, a ``bow tie'' diagram within the distributed systems described above, is a descriptive mechanism for broadcasting control signals to multiple recipients. The Global Neuronal Workspace, as a massively parallel, distributed system, performs precisely this function for the mammalian central nervous system \cite{baars:03,dehaene:01,mashour:20}.

We refer the reader to the concepts of a (regular) {\em theory} and {\em local logic} as introduced in \cite[Chs 9~12]{barwise:97} (and extensively reviewed in \cite{fg:19a,fg:21}) to specify the logical structure of a given situation. Basically, a local logic is a classifier with a theory, along with a subset of tokens as specified by a {\em sequent}; namely, a classification $M \models_{\mathcal{A}} N$ of a classifier given by a pair of subsets $M,N$ of $Typ(\mathcal{A})$, such that $\forall x \in Tok(\mathcal{A}), x \models_{\mathcal{A}} M \Rightarrow x \models_{\mathcal{A}} N$.  In fact, any classifier generates its own natural local logic.  Below, we will adopt the sequent as a `conditional' when defining probabilities.

\subsection{CCCDs implement Bayesian inferences in defined contexts} \label{Bayesian}

In general, given an information flow channel:

\begin{equation}\label{channel-flow-}
\longrightarrow \mathcal{A}_{\alpha -1} \longrightarrow \mathcal{A}_{\alpha} \longrightarrow \mathcal{A}_{\alpha +1} \longrightarrow \cdots
\end{equation}
\noindent
the semantic content can be extended by postulating local logics $\mathcal{L}_{\alpha} = \mathcal{L}(\mathcal{A}_{\alpha})$ generated by the corresponding classifiers $\mathcal{A}_{\alpha}$ (assumed, in principle, to be in relationship to a (regular) theory associated to the individual $\mathcal{A}_{\alpha}$, as specified in \cite[Ch 9]{barwise:97} and reviewed in \cite[Appendix A]{fg:21}), thus postulating a flow of logic infomorphisms:

\begin{equation}\label{cbd-flow-2}
\cdots \longrightarrow \mathcal{L}_{\alpha -1} \longrightarrow \mathcal{L}_{\alpha} \longrightarrow \mathcal{L}_{\alpha +1} \longrightarrow \cdots
\end{equation}
\noindent
which may then comprise a CCD as in Eq. \eqref{ccd-1}.

A probabilistic interpretation of information flow in Eq. \eqref{cbd-flow-2} results when the sequent relation is weakened to require only that if $x \models_{\mathcal{A}} M$, there is some probability $P(N|M)$ that $x \models_{\mathcal{A}} N$. \footnote{We use the upper case `P' to denote a probability in relationship to classifiers, and a lower case `p' for that pertaining to e.g. events or states as below.} In effect, $\models_{\mathcal{A}}$ becomes a conditional probability that inherits the semantics of the local logic \cite{allwein:04a} (cf. \cite{barwise:97a}); that is, given a sequent in a classifier $\mathcal{A}$, along with its satisfaction relation satisfying:

\begin{equation}\label{prob-1}
M \models_{\mathcal{A}} N ~, \qquad \forall x (x \models_{\mathcal{A}} M \Rightarrow x \models_{\mathcal{A}} N),
\end{equation}
\noindent
we can define the conditional probability as:

\begin{equation}\label{prob-2}
M \models_{\mathcal{A}}^{P} N := P( M \vert N).
\end{equation}
\noindent
In fundamental Bayesian terms, $M$ above can be regarded as an unobserved (or as below, unobservable) event, and $N$ an observable datum, in which case $P(M)$ becomes the prior, and $P(N)$ the evidence, that together generate a prediction. Given the likelihood $P(N \vert M)$ as the conditional obtained from weakening the sequent, Bayes' theorem specifies this conditional as the posterior:

\begin{equation}\label{prob-3}
P(M \vert N) = \frac{P(N\vert M) P(M)}{P(N)}.
\end{equation}

We introduce the idea of ``contexts'' of observation, following \cite[\S 7.1]{fg:21} (see also \cite[Appendix A.1]{fgm:21}), by considering the following sets: i) $X$ is a set of events in a very general sense (e.g. a set of observed value combinations or atomic events); ii) $Y$ is a set of conditions (specifying objects/contents or influences; iii) $Z$ is a set of contexts (e.g. detectors, measurements, or methods); iv) $W := Y \times Z$, where $X, Y$ and $Z$ are taken to be subsets of some (very) large probability space.  Note that here $Y$ can be decomposed as $Y = Y^{+} \cup Y^{-}$ (disjoint union) where $Y^{+}$ consists of observed objects/contents/characteristics of the context $Y$, and $Y^{-}$ consists of what is not observed about $Y$. Hence $W: = Y \times Z = (Y^{+} \cup Y^{-}) \times Z$. The classifier in question is then $\mathcal{A} = \langle X, W, \models_{\mathcal{A}} \rangle$ and represents an {\em observable in context}. Among several possible interpretations for the classification relation `$\models_{\mathcal{A}}$' is valuation by the conditional probability $p(a\vert x) = p(a\vert \{b,c\})$, whenever defined, for $a \in X, b \in Y$ and $c \in Z$ \cite{fg:21} (see also below).

As a brief example, consider arbitrary classifiers $\mathcal{A}_1^{(a)}, \ldots, \mathcal{A}_5^{(e)}$ in some part of an information channel where the classifiers correspond to events $a,b,c,d,e$, respectively, together with logic infomorphisms $f_{13}, \ldots, f_{45}$ between them, and where the sequents are relaxed to become conditional probabilities as above:

\begin{equation}
\begin{gathered}
\xymatrix{&\mathcal{A}_1^{(a)} \ar[dl]_{f_{13}} \ar[dr]^{f_{14}} &  & \mathcal{A}_2^{(b)}\ar[dl]_{f_{24}} \\
\mathcal{A}_3^{(c)} & & \mathcal{A}_4^{(d)} \ar[d]^{f_{45}}  &  & \\
& & \mathcal{A}_5^{(e)}
}
\end{gathered}
\end{equation}
\noindent
Following e.g. \cite{cherniak:91}, this particular channel then generates a joint probability distribution given by:

\begin{equation}
p(abcde) = p(a)p(b)p(c\vert a)p(d \vert ab)p(e\vert d).
\end{equation}

As developed in detail in \cite{fg:21}, the above constructions provide a formal basis for classical Bayesian inference. Specifically, the diagrams \eqref{ccd-1} and \eqref{cccd-1} depict channels of logic infomorphisms, when on relaxing the sequent in each case, the classifications $\models_{\mathcal{A}_{\alpha}}$ are now explicitly realized as conditional probabilities $p_{\alpha}( \cdot \vert \cdot)$, whenever these are defined.  Putting the above details concerning conditionals into the diagrammatic framework reveals a typical portion of a CCD computing a hierarchical Bayesian inference from a set of (posterior) observations $\mathcal{A}_i$ to an outcome $\mathbf{C}'$, as having the form:
\begin{equation}\label{ccd-prob-1}
\begin{gathered}
\xymatrix@C=4pc{&\mathbf{C^\prime} &  \\
\mathcal{A}_1 \ar[ur]^{p_{10}(\cdot \vert \cdot)} \ar[r]_{p_{12}(\cdot \vert \cdot)}& \mathcal{A}_2 \ar[u]_{p_{20}(\cdot \vert \cdot)} \ar[r]_{p_{23}(\cdot \vert \cdot)} & \ldots ~\mathcal{A}_k \ar[ul]_{p_{k0}(\cdot \vert \cdot)}
}
\end{gathered}
\end{equation}
The requirement that diagrams \eqref{ccd-1} and \eqref{ccd-prob-1} commute amounts to the requirement that branching ``upward'' to an overall logic $\mathcal{L}$ along any one of the $f_{\alpha}$, is equivalent to following the inferential sequence \eqref{cbd-flow-2} up to its termination, and then following the last of the $f_\alpha$ to $\mathcal{L}$.  Thus the $f_{\alpha}$ can be seen as shortcuts to reaching $\mathcal{L}$: ``insights'' that allow the rest of the inferences in \eqref{cbd-flow-2} to be bypassed.  The local logic $\mathcal{L}$ can be seen as the ``answer'' to the problem \eqref{cbd-flow-2} addresses: formally, it is the logic that solves the problem in one step.  The probability that the answer is ``right'' is the product of the probabilities along \eqref{cbd-flow-2}.  The commutativity requirement on \eqref{ccd-prob-1} requires, in addition, that this overall probability is conserved; hence the probability associated with each ``insight'' $f_{\alpha}$ must be the combined probability of the inferential steps it replaces.

 Commutativity of a CCD thus enforces inferential coherence, and the same clearly applies when the dual objects and maps of a CD are combined in constructing a CCCD as in diagram \eqref{cccd-1} to implement a recurrent Bayesian network (see also \cite{fg:19b} where CCCDs are employed to model the interaction between bottom-up and top-down processing in visual object categorization).  In contrast, non-commutativity of \eqref{ccd-1} implies that there is no consistently definable (conditional) probability distribution across the system in question, and characterises {\em intrinsic contextuality} \cite{kochen:67, mermin:93, dzh:18} for that system \cite[Th 7.1 and Coroll 7.1]{fg:21}.

\subsection{QRFs as memory resources} \label{attractors}

A QRF is only useful operationally, and hence only meaningful as a reference frame, if it can be deployed consistently over multiple episodes of observation.  A meter stick or a voltmeter is, clearly, useful only if yields consistent measurements; the notion of a ``standard'' to which a device can be periodically recalibrated captures this consistency requirement operationally.

Recalling that a QRF is by definition a physical system, not merely an abstraction \cite{bartlett:07}, the above condition can be stated as a requirement that a QRF have stable dynamics over the relevant time period, or if it is a subsystem of a larger system $A$, that it is an (or a component of an) attractor of the dynamics of $A$, i.e. of the self-interaction $H_A$.  In computational language, a QRF must be a memory structure, one that is active (and hence executed as a computation) at all times, or one that is executable on demand via some higher-level control structure.  We can interpret a CCD identifying $S = RP$ that is implemented by $A$ as encoding an expectation, and hence a prior probability for $A$, that bit patterns specifying $S$ will be encoded on the holographic screen $\mathscr{B}$ separating $A$ from its complement $B$.  As emphasized earlier, neither the existence of this expectation not its satisfaction in practice imply that $S$ exists in any ``ontic'' sense as an observer-independent component of $B$ (cf. \cite{prakash:20}).

\begin{figure}[H]
\centering
\includegraphics[width=13 cm]{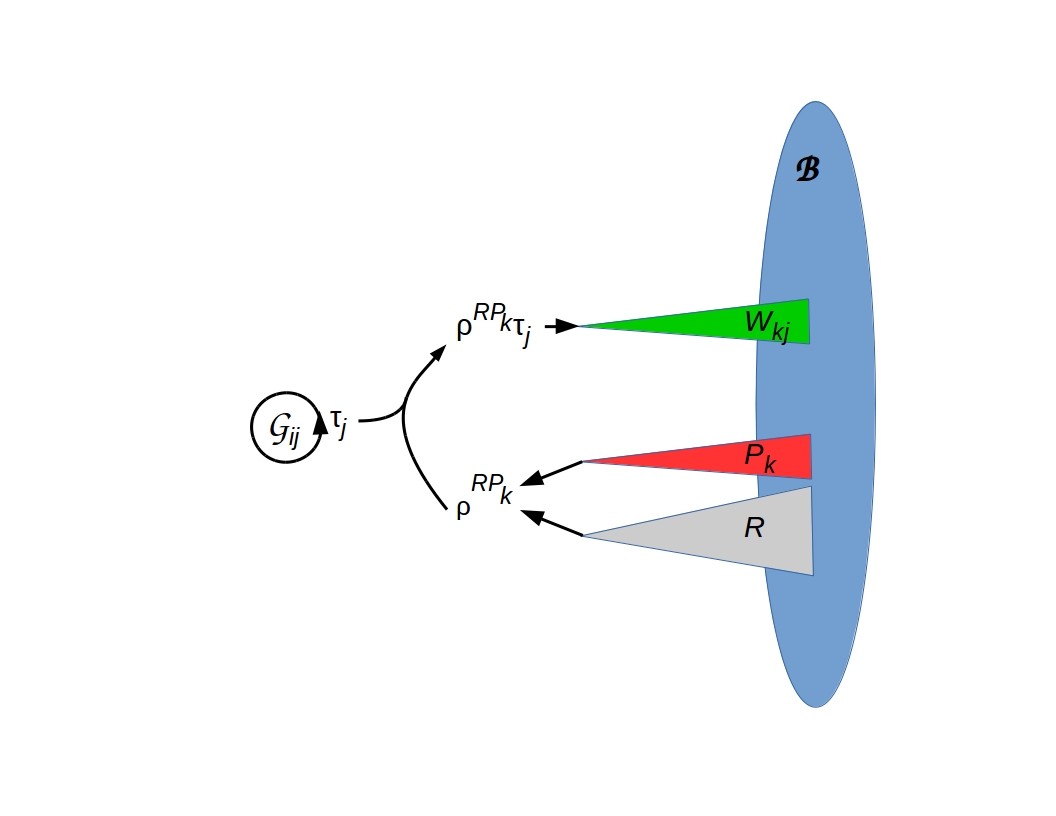}
\caption{A CD $W_{kj}$ (green triangle) specifies a memory-write operation of the time-stamped state $(\rho^{RP_k}, \tau_j)$ to $\mathscr{B}$.  The timestamp $\tau_j$ is generated by the groupoid action $\mathcal{G}_{ij}$ on the previous (at $\tau_i$) output from the CCD $PR$.  Adapted from \cite{fgm:21} Fig. 9, CC-BY license.}
\label{QRF-mem-write-fig}
\end{figure}

Operational utility and hence meaningfulness also require that a QRF write its outcomes to a classical memory in a way that allows comparison of outcomes obtained at different times.  Following \cite{fg:20a, fgm:21}, we consider the transition from writing an observed pointer outcome $P_i$ with timestamp $\tau_i$ to writing $P_k$ with timestamp $\tau_j$ to be implemented by an operator $\mathcal{G}_{ij}$, the complete set of which forms a groupoid under composition.  ``Timestamping'' here can be considered merely to be imposition of an order on the recorded data; a separate, observation independent clock is not assumed (cf. \cite{fgl:21} for a general discussion of time QRFs).  As compliance with Eq. \eqref{ham} requires all classical data to be written on the holographic screen $\mathscr{B}$, the memory-write CD must write on $\mathscr{B}$ as shown in Fig. \ref{QRF-mem-write-fig}.  Comparing previous with current data, in this case, requires reversing the arrows of the relevant memory-write CD to reconstruct the previously-written data.  The comparison itself becomes a ``meta'' operation, i.e. one defined at the next-higher hierarchical level as described in the case of neurons below (cf. \cite{fg:19b, fg:20, fg:21} for examples from human cognitive processing).

We can, therefore, consider QRFs to provide two distinct memory resources: 1) the stable QRF itself as an executable computation, and 2) the ordered sequence of classical data accumulated by deploying the QRF to make measurements.  The executable QRF is, in general, a ``quantum'' memory, as it is a quantum computation implemented by the operator $H_A$.  The classical data, in contrast, are written on $\mathscr{B}$ and subject to both space and free-energy constraints.  As discussed in \cite{fgl:21}, these latter memories are effectively stigmergic.  Any representation of the ``self'' that is readable as a classical memory, in particular, must be written on $\mathscr{B}$, and read from $\mathscr{B}$ when recalled \cite{fgl:21}.  These distinct memory resources support distinct expectations: QRFs as executable computations encode expectations about how the ``world'' faced by an agent is (or more properly, usefully can be) decomposed into identifiable ``systems'' while the classical data encode expectations about the states in which these systems will be encountered.  These expectations can be represented as Markov kernels that are learned incrementally; the difference between these ``expected'' kernels and the ``true'' kernels given by Diagram \eqref{diag1} provides a representation of VFE to be minimized by active inference, i.e. by exploration (novelty seeking that enlarges the ``training set'') and further learning.  This definition of VFE for generic quantum systems allows a formulation of the FEP \cite{friston:10a,friston:13a,friston:07a,ramstead:20,friston:19} for such systems that is developed in detail elsewhere \cite{ffgl:21}.

\section{QRFs at the macromolecular scale} \label{macromolecular}

We now turn from generalities to the specific case of mammalian cortical neurons, and show how these can be conceptualized as hierarchies of QRFs, i.e. hierarchies of physically-implemented computations, each of which can be regarded as a measurement, at a defined scale and with respect to defined units, of some aspect of its local environment.  We then generalize this treatment to arbitrary cells in \S\ref{cells}.

\subsection{QRFs as homeostatic setpoints}

As discussed in \cite{fl:20, fgl:21}, the simplest biological QRFs are switches that induce opposing behaviors whenever some parameter value is above or below some threshold value.  The CheY system regulating bacterial chemotaxis provides a familiar example: local concentrations of the phosphorylated form CheY-P above or below a essentially fixed (at the relevant timescales) default concentration induce swimming or tumbling behavior, respectively \cite{lyon:15, micali:16}.  From a computational perspective, the CheY system implements a switch between depth-first and breadth-first search, and hence a generic mechanism for avoiding local maxima in a search space.  More biologically, it implements both approach to resources and avoidance of toxins and other irritants, and hence serves as a homeostatic setpoint. Such setpoints are, from the present perspective, expectations consistent with continuing structural and functional integrity, i.e. continuing separability from the surrounding world.  They can, therefore, be expected to arise in any system that maintains separability, i.e. any system to which Eq. \eqref{ham} applies.

A second familiar QRF is the default or ``resting'' membrane voltage $V^0_{mem}$ across any local patch of biological membrane.  Fluctuations above or below this setpoint induce actions, e.g. opening or closing voltage-gated ion channels.  This QRF is obviously relevant to neuron function as discussed below; at larger scales, it becomes a critical regulator of multicellular morphology; see e.g. \cite{levin:12, levin:17, levin:21a} and further discussion in \S\ref{cells}.

\begin{figure}[H]
\centering
\includegraphics[width=15 cm]{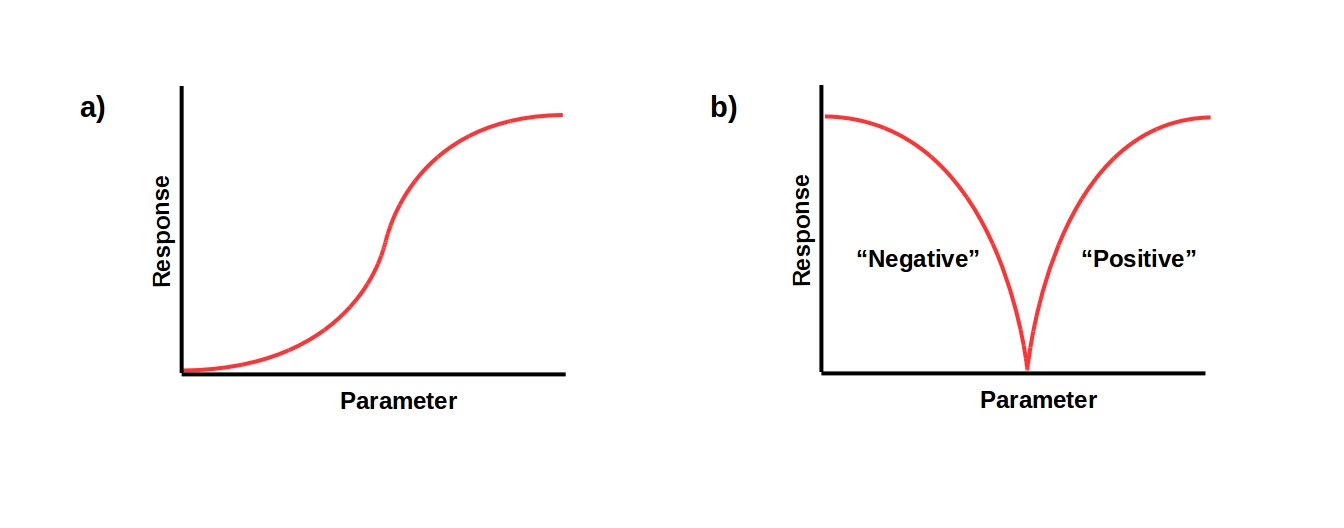}
\caption{a) A generic system executing sigmoid kinetics.  b) Reflecting the response curve at its inflection point redescribes the system as a QRF that switches behavior at a default value.}
\label{sigmoid-response-fig}
\end{figure}

Guided by these examples, it is straightforward to consider any system implementing linear threshhold or sigmoid kinetics as a QRF that switches between ``negative'' and ``positive'' responses as its input fluctuates below or above a default value (Fig. \ref{sigmoid-response-fig}).  With the advent of two-component signaling pathways, early enough in prokaryotic evolution that they appear in all analyzed lineages \cite{wuichet:10}, the default value becomes adjustable, i.e. context-dependent via a classical ``direct influence'' mechanism \cite{dzh:18} (but see \cite{basieva:11} for evidence that lactose-glucose interference in {\em E. coli} \cite{inada:96} exhibits nonclassical contextuality).  The multicomponent systems typical of eukaryotes, e.g. the Wnt \cite{loh:16}, ERK \cite{kolch:05}, notch \cite{schwanbeck:11}, BMP \cite{guo:09} pathways and the interactions between them, introduce further, multifactorial context dependence (see e.g. \cite{hunter:00, adamska:15} for general reviews).  Such systems embody expectations about the relationships between multiple, simultaneously-measured values, each effectively calibrated to an underlying standard, e.g. a molecular concentration or local value of $V_{mem}$.

As noted earlier, cellular-scale biochemical and bioelectric signaling pathways have traditionally been conceptualized and depicted as fully classical, with each component occupying some determinate state at all relevant times.  Cellular energy budgets cannot, however, support the thermodynamic cost of classicality, even at ms timescales, indicating that quantum coherence and hence quantum computation may be significant up to timescales of seconds \cite{fl:21}.  Classical data can, in this case, only be encoded on intra- or intercellular boundaries as illustrated schematically in Fig. 1.  ``Chunking'' the cell into functional components that process and exchange classical information or serve as classical memory structures (e.g. the genome, proteome, transcriptome, and ``architectome'' \cite{fl:18}) is, effectively, identifying the boundaries at which classical information is encoded, i.e. the boundaries that function as MBs.  A signal transduction pathway that measures some environmental parameter at the cell membrane and transfers a context-dependent representation of this information to the genome is, in this picture, a single quantum computation, a QRF that detects, calibrates, and reports an observational outcome.  It can be treated as a black box characterized by inputs and outputs, including free energy inputs and dissipated heat outputs.  We adopt this approach to characterize the functional architecture of neurons in what follows.

\subsection{The postsynaptic complex as a QRF}

\begin{figure}[H]
\centering
\includegraphics[width=15 cm]{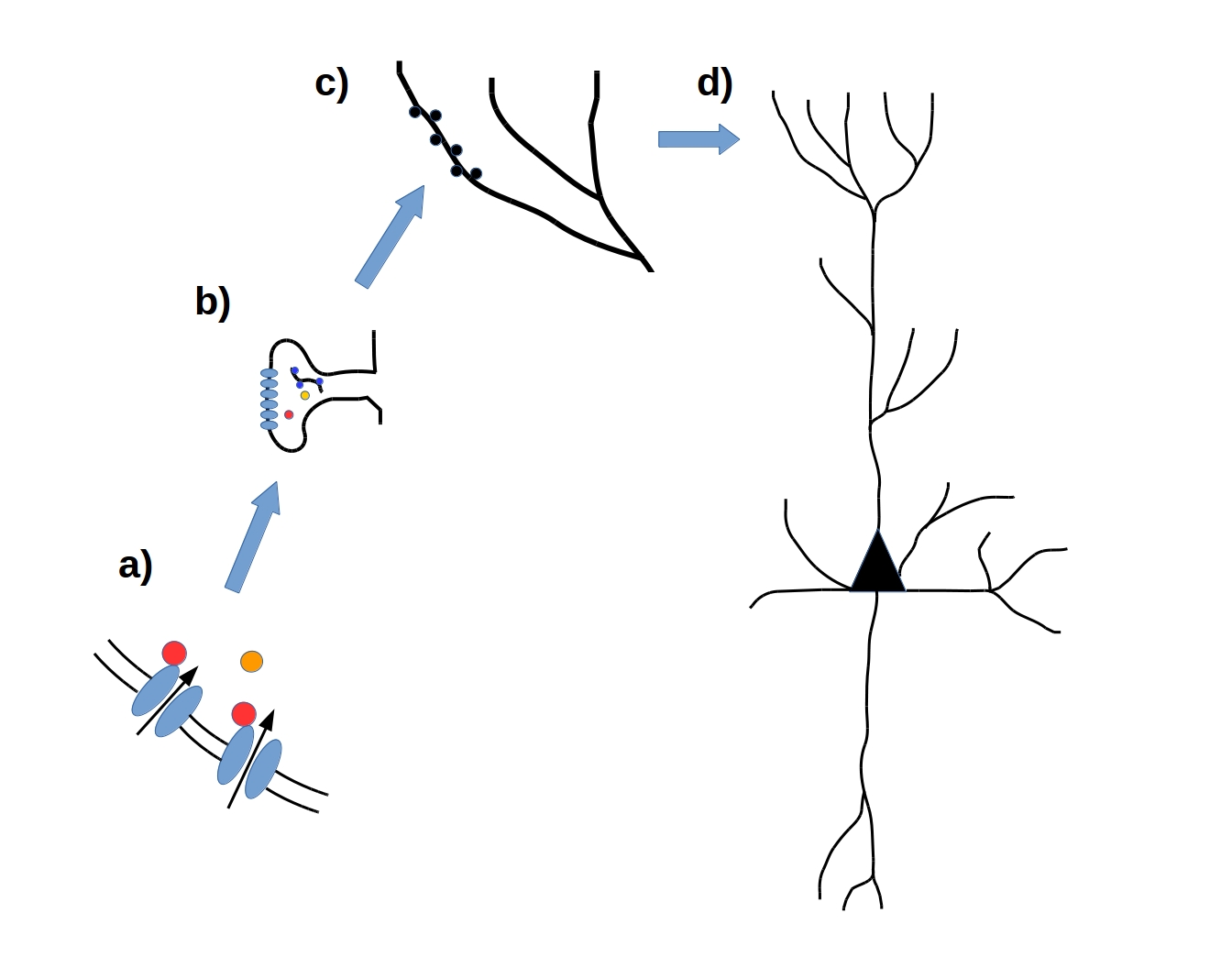}
\caption{Cartoon structure of a mammalian cortical neuron; cf. \cite{spruston:08, rasia:21} for anatomical details.  a) Voltage gated channels and pumps are the primary input (sensory) and output (active) MB components; b) Specialized postsynaptic structures, e.g. on dendritic spines, collect incoming information from other neurons; c) dendritic subtrees integrate and process this incoming information; d) dendritic information is integrated at the soma and distributed to outgoing axonal channels.  Postsynaptic receptors may also appear on the soma or on axons; this level of detail is neglected here.}
\label{neuron-components-fig}
\end{figure}

As discussed in \cite{hipolito:21,palacios:20,peters:17}, the MB of a single neuron includes the pre- and postsynaptic membranes that mediate interactions with other neurons, as well as the overall cell membrane that mediates interactions with the neuron's local microenvironment.  Single neurons are, however, large, spatially-extended, computationally-complex structures (Fig. \ref{neuron-components-fig}); it is the need for improved models of such structures that motivates this paper.  Hence we will consider how the neuron's MB is constructed from those of its components, identifying in this process sites at which intermediate steps in neuronal computations may be classically encoded.

\begin{figure}[H]
\centering
\includegraphics[width=15 cm]{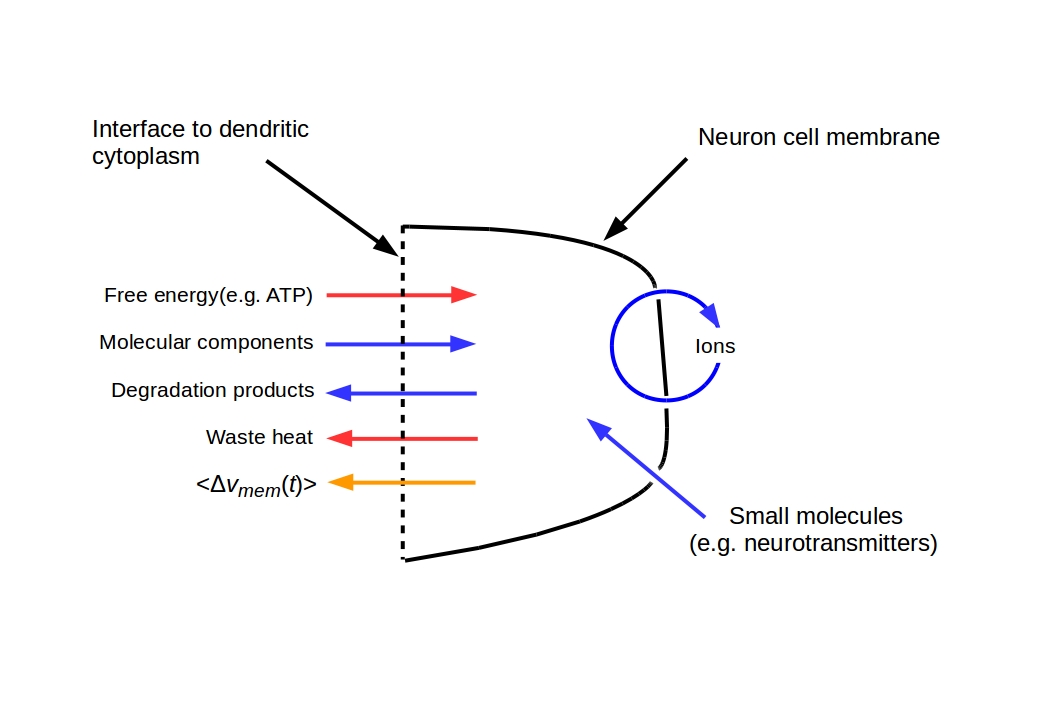}
\caption{Cartoon structure of a mammalian postsynaptic specialization; cf \cite{runge:20} for anatomical and functional details.  Primary energy (red), molecular (blue) and $V_{mem}$ gradient (orange) flows are shown.  Considered as a system, the spine's MB includes the spine membrane (solid line) and the cytoplasmic/cytoskeletal interface between the spine and the bulk dendritic cytoplasm (dashed line).}
\label{PS-specialization-fig}
\end{figure}

Starting from the input side, the smallest cellular-scale component is the specialized postsynaptic area, typically located on a spine morphologically.  Its computational role is to contribute a positive (excitatory) or negative (inhibitory) time-dependent $V_{mem}$ gradient to the overall electrical activity of its local dendritic branch.  Performing this function requires energetic and molecular flows in addition to $\Delta V_{mem} (t)$, as illustrated in Fig. \ref{PS-specialization-fig}.  At this level of abstraction, the presynaptic specialization at the axonal bouton is essentially equivalent, up to reversing the flow of small molecules from the cell membrane.

\begin{figure}[H]
\centering
\includegraphics[width=15 cm]{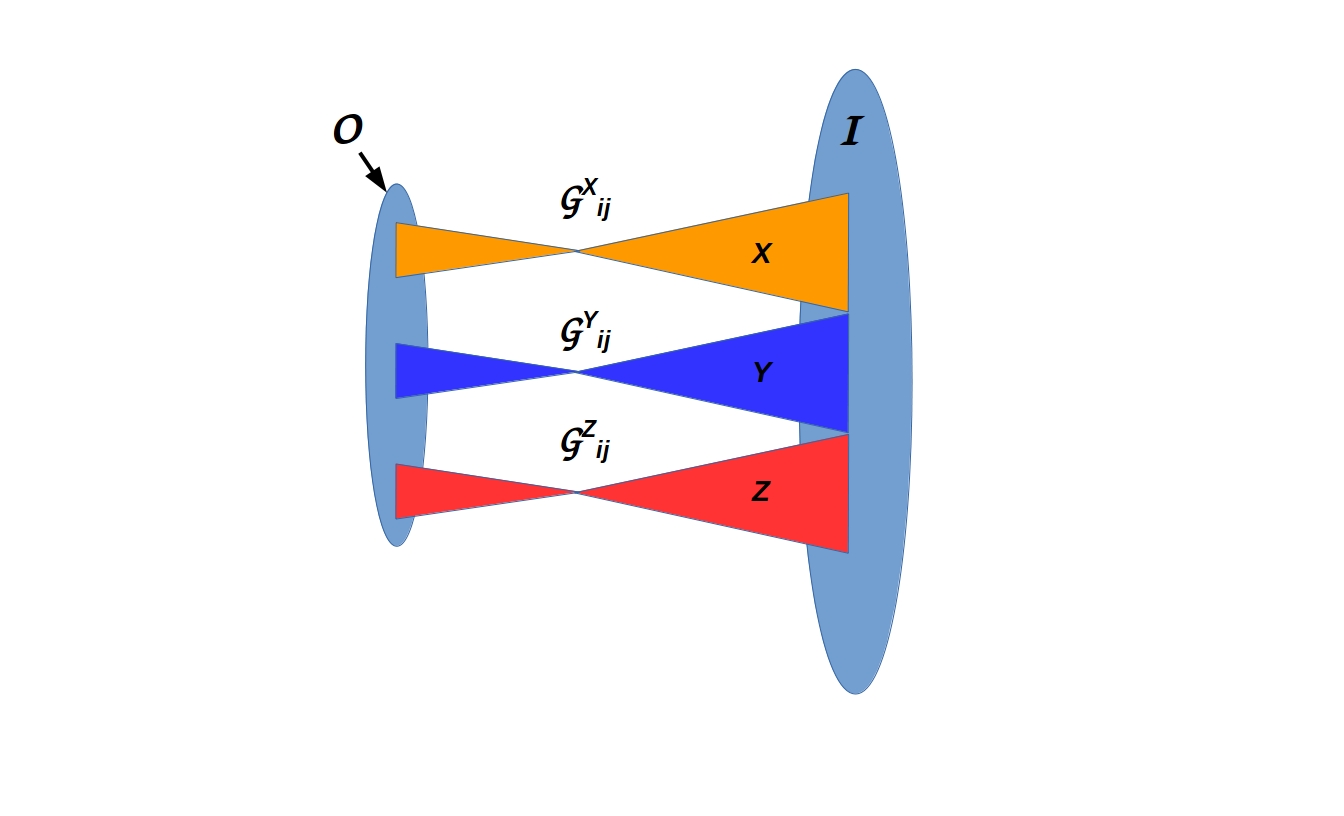}
\caption{Information flows abstracted to QRFs ($X$, $Y$, $Z$; colored triangles) that read from an input boundary component $\mathcal{I}$ (right hand blue ellipse) and write to a output boundary component $\mathcal{O}$ (left hand blue ellipse), where the overall system boundary $\mathscr{B} = \mathcal{I} \mathcal{O}$.  Each QRF $Q$ has an associated time constant $\tau^Q$ and a groupoid-structured set of operators $\mathcal{G}^Q_{ij}$ as depicted in Fig. \ref{QRF-mem-write-fig}.}
\label{QRF-flow-fig}
\end{figure}

Making the reasonable assumption of balanced mass flows, i.e. assuming that the volume and density of the postsynaptic specialization, or in quantum terms, its Hilbert-space dimension remains constant, we can focus on information flows only, and treat the interaction between the postsynaptic specialization and its surroundings using Eq. \eqref{ham}.  The boundary $\mathscr{B}$ in this case has two components, $\mathscr{B} = \mathcal{C} \mathcal{D}$, where $\mathcal{C}$ separates the postsynaptic specialization from the external microenvironment and $\mathcal{D}$ separates it from the bulk dendritic cytoplasm.  Both components support both inward sensory flows and outward active flows.  As these information flows are both physically implemented and referenced to default homeostatic/allostatic setpoints, they can be considered to be implemented by QRFs.  Hence they can be represented as input -- output or perception -- action flows as in Fig. \ref{QRF-flow-fig}.  In this representation, the boundary $\mathscr{B}$ is decomposed functionally into $\mathscr{B} = \mathcal{I} \mathcal{O}$, where $\mathcal{I}$ and $\mathcal{O}$ are ``input'' and ``output'' components respectively, each of which includes regions of both anatomical components $\mathcal{C}$ and $\mathcal{D}$.  While each flow $Q$ has its own characteristic time constant $\tau^Q$ as in Fig. \ref{QRF-mem-write-fig}, the cross-modulatory couplings between pathways, and hence mutual dependence between setpoints, requires that they be mutually coherent.  Hence we can represent the postsynaptic specialization as executing a single QRF with a vector output and a time constant $\tau^{PS} = max_Q \tau^Q $.  Presynaptic specializations can be treated similarly.

As discussed in \S\ref{QRFs} above, all classical information in the system must be encoded on $\mathscr{B}$.  Hence Fig. \ref{QRF-flow-fig} depicts the ideal case in which the boundary component $\mathcal{O}$ encodes {\em all} output information, i.e. the computations implemented by the QRFs $X$, $Y$, and $Z$ have no classically-encoded intermediate states.  In this ideal case, the free-energy cost of computation per unit time is given by the output bandwidth in bits, i.e. the area in bits of $\mathcal{O}$.  Assuming a minimal thermal energy $\Delta E_{th} = \mathrm{ln2}k_B T$, the minimum dissipation timescale is given by the time-energy uncertainty relation \cite{lloyd:00} as $\Delta t_{diss} \geq \pi \hbar / (2 \Delta E_{th} ) \approx$ 50 fs.  Neuronal energy budgets, however, are insufficient for classical encoding at this timescale; assuming 30,000 synapses per cortical neuron \cite{rasia:21} and a maximum power consumption of 250 Gbits/sec (where as a power unit 1 bit $=_{def} \mathrm{ln2}k_B T \approx 3 \cdot 10^{-21}$ J at $T = 310$ K) \cite{herculano:11}, each synapse could encode at most 8.3 Mbits/sec if the cell's metabolic resources were devoted entirely to synaptic encoding.  In fact the bulk of neuron energy usage is devoted to action potential (AP) generation and transmission, reducing the coding capacity of synapses by up to two orders of magnitude (see \cite{fl:21, attwell:01, sengupta:13, georgiev:20a} for further discussion of neuronal classical encoding costs and capacity).  We can expect, therefore, that individual postsynaptic specializations can realistically encode at most about 100 kbits/sec, or 100 bits per unit time at the ms scales of dendritic postsynaptic potentials.  Hence the idealization of Fig. \ref{QRF-flow-fig} is not unrealistic from an energetic perspective, suggesting that consistent with the general considerations in \cite{fl:21}, QRFs implemented at the scale of cellular compartments perform essentially pure quantum computations, i.e. employ quantum coherence as a short-term memory resource.

The form of Fig. \ref{QRF-flow-fig} suggests a cobordism with $\mathcal{I}$ and $\mathcal{O}$ as boundaries.  If the QRFs $X$, $Y$, and $Z$ perform pure quantum computation without classical intermediate steps, we can treat them in a path-integral formalism \cite{deutsch:02}, and hence as defining a manifold of mappings from $\mathcal{I}$ to $\mathcal{O}$.  This suggests that information processing in neurons is amenable to a treatment using topological quantum field theory \cite{atiyah:88} and the emerging theory of topological quantum neural networks \cite{marciano:21}.  We defer this possibility to future work.

\section{Dendrites as spatially-organized QRFs} \label{dendrites}

Dendritic branches are traditionally viewed as lossy but otherwise passive integrators of relatively slow (10s of ms) postsynaptic potentials (PSPs), though faster spike-like activations and backwards propagation of APs are also possible \cite{spruston:08}; see \cite{eyal:18} for detailed dendritic spike models.  This passive view of dendritic activity renders the execution of complex logical operations like XOR \cite{gidon:20} surprising and activity-dependent remodeling \cite{carulli:11, hogan:20, runge:20} mysterious.  It decouples the processes required to maintain a successful synapse: learning is largely localized to the postsynaptic side, while target search and synapse formation are localized to the presynaptic side.  Finally, it decouples dendritic function from the dendrite's trophic requirements.  Branches that transmit only low-amplitude noise signals cannot, on the passive model, be regarded as less successful or less worthy of continued metabolic maintenance in comparison to those branches that consistently transmit high-amplitude, highly informative signals.

Dendritic branches satisfy the physical requirements for implementing QRFs and hence engaging in active inference.  Viewing dendrites in this way raises two immediate questions:

\begin{enumerate}
\item What is a dendritic branch measuring?
\item What is the VFE function that a dendritic branch is minimizing?
\end{enumerate}
\noindent
However, while dendrites can be described as coincidence and hence activity-correlation detectors \cite{spruston:08} even on the passive view, the second of these questions is difficult to even formulate from a passive perspective.  A partial answer to both questions is discussed in \cite{kiebel:11, bastos:12}: basically, dendrites can self-organize to minimize the VFE on surprise of their presynaptic inputs, showing that postsynaptic gain is itself optimized with respect to VFE.  Extending the modeling approach of the previous section to dendritic branches suggests answers to these questions that can be summarized by the hypothesis that dendritic branches identify ``objects'' with relatively small numbers of states, and execute approximately binary logic on the measured state vectors.

\subsection{Colocating consistently correlated synapses minimizes VFE}

While both dendritic arborization and spine density vary with neuron type and location \cite{galloni:20,major:13,shipp:07,shipp:16}, both cortical and hippocampal pyramidal cells can exhibit dendritic branches with on the order of 1,000 spines and full dendritic trees with up to 30,000 spines \cite{rasia:21}.  Consider such a branch, neglecting information and resource flows through the non-spine membrane to focus on flows to and from the spines as modeled in Fig. \ref{PS-specialization-fig}.  As the magnitudes of resource flows correlate with activity and hence with the magnitude of $\Delta V_{mem}$, we can further focus on the latter.  In this case, we can think of the branch as ``observing'' on the order of 1,000 points of active signaling, analogous to 1,000 points of light in a visual field.

There are clearly two limiting cases for the activity observed by a branch.  If every spine receives input from a distinct, independently-activated neuron, the correlation $C_{ij}$ between inputs $i$ and $j$ can be expected to be small, with $C_{ij} \rightarrow 0$ in the limit.  If all spines receive input from a single neuron, $C_{ij} \rightarrow 1$ in the limit.  Assuming roughly constant frequencies (though random phases) of presynaptic activity, the former limit yields noise with an input-frequency dependent amplitude as an output PSP from the branch; the latter yields an essentially digital signal encoding the activity of the single input neuron.  An essentially constant noise signal is effectively a ``dark room'' from a VFE perspective, while a variable digital signal poses a well-defined, potentially high-information prediction problem.

Microconnectome methods \cite{swanson:16} allow, in principle, counting the numbers of presynaptic (i.e. incoming) and postsynaptic (outgoing) partners for any neuron.  Measured numbers of presynaptic partners range up to 50 in {\em C. elegans} \cite{varshney:11} and from tens to several hundred in mammalian cortical cells \cite{velez-fort:14, denardo:15} to thousands for cerebellar Purkinje cells \cite{swanson:16}.  If as estimated \cite{harris:15} neocortical pyramidal cells typically receive 4 or 5 synaptic connections from each presynaptic partner, these cells may also have several thousand presynaptic partners.  Such cells would be operating very close to the noise limit of $C_{ij} \rightarrow 0$ unless the activity of the presynaptic partners is already highly correlated.  Perhaps significantly, most microscale connectome mapping has been achieved in sensory cortices in which highly correlated incoming information can be expected.

\begin{figure}[H]
\centering
\includegraphics[width=15 cm]{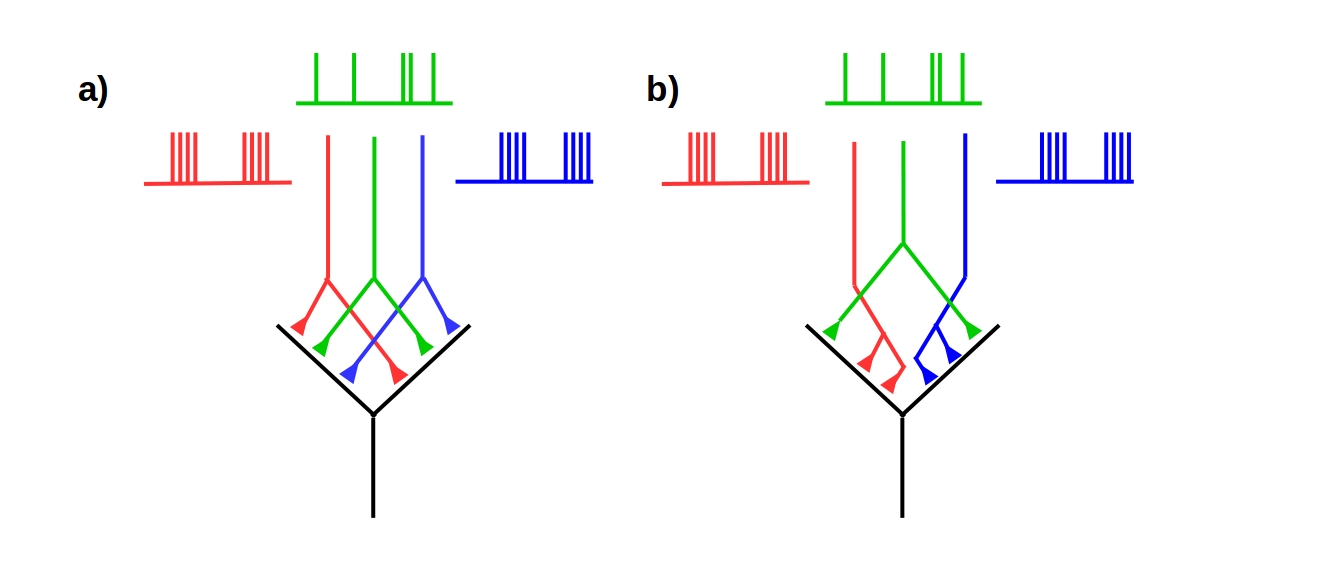}
\caption{a) Least and b) most efficient arrangements of synapses with partially-correlated (red and blue) and random (green) activity on a two-sided symmetrical, terminal dendritic branch.  By minimizing the ambiguity of correlation, arrangement b) allows maximally digital processing at the branch junction.}
\label{incoming-correlates-fig}
\end{figure}

Following \cite{spruston:08} and considering each dendritic branch to be a coincidence detector, the fundamental problem faced by a branch is distinguishing between i) correlated signals from a given presynaptic partner, ii) ``true'' correlates from multiple partners, and iii) ``random'' correlates.  We can treat this uncertainty as a VFE function, and ask how it can be minimized.  For a relatively short, distal, terminal branch exhibiting all three kinds of correlates, the answer is shown in Fig. \ref{incoming-correlates-fig}: localizing partially-correlated regular inputs to distinct, non-branched branches and moving random inputs as far as possible from branch junctions disambiguates signals from different synaptic partners, allows ``true'' correlations to be identified specifically at branch junctions, and reduces the amplitude of noise signals.

A terminal branch in an initial state similar to Fig. \ref{incoming-correlates-fig}a has synapse-level learning and spine remodeling available as tools for moving toward a state similar to Fig. \ref{incoming-correlates-fig}b, with trophic reward as the imposed quality function.  This is active inference, with spine remodeling as the ``action'' that alters environmental input, i.e. synaptic activity, independently of synapse-level priors.  Success (given enough time) can be expected if trophic reward increases with the overall informativeness of the branch's activity, i.e. if trophic reward to dendrites as functional units has the same activity-dependence as trophic reward to neurons as functional units \cite{butz:09, mennerick:00, faust:21}, as studies of dendritic remodeling already suggest \cite{hogan:20, runge:20}.  This is mechanistically reasonable, as it transfers both the task of determining informativeness and that of adjusting trophic reward to the proximal part of the dendritic tree, and hence effectively to the soma itself.  Both are, from the branch's perspective, outside of its MB and hence part of the environment.

We can, therefore, state the following:

\begin{quote}
{\bf Prediction:} Trophic reward to dendritic branches correlates with informativeness of the signal from the branch for more proximal dendritic branches, with informativeness to the soma as a limit.
\end{quote}

\noindent
What the soma needs to know is whether to fire an AP.  A high-amplitude signal on a low-amplitude background, i.e. an effectively digital PSP, provides this information most efficiently.  Hence we can predict that dendritic trees will remodel in the direction of maximizing digital processing.

\subsection{Branches as object detectors}

A branch that has segregated its non-noise inputs into different compartments as in Fig. \ref{incoming-correlates-fig}b is, effectively, an object detector.  The object it detects is a presynaptic partner, or a spatially-localized collection of axonal processes from a presynaptic partner.  It measures the detectable state of this object, its binned-average activity at some binning timescale, typically a few ms.  Hence we can think of it on the model of Eq. \eqref{diag1} or Fig. \ref{QRF-mem-write-fig}, with the input correlation as the reference $R$, the binned-average activity as the pointer $P$, and any background noise as the environment $E$.  Hence a compartmentalized branch is a canonical collection of QRFs.

From this perspective, we can see spine remodeling within dendritic branches as a process of assembling QRFs that identify objects and hence enable measurements of their states.  Spine remodeling is thus an instance of the second, relatively neglected, form of active inference: the minimization of uncertainty about the decomposition of an agent's world.  A branch with maximally-ambiguous inputs as in Fig. \ref{incoming-correlates-fig}a decomposes its world only into ``points of light''; a branch that has remodeling to approach Fig. \ref{incoming-correlates-fig}b sees instead two objects with well-defined states that can be compared in both amplitude and time dimensions.  Spine remodeling is, as are the processes of object segregation and object persistence during infant development, learning {\em what} to see, and hence learning what coherent things populate the perceived world.

\section{Neurons as measurement devices} \label{neurons}

\subsection{Dendritic geometry imposes salience}

As the ``cut'' and hence the boundary imposed to render a branch ``terminal'' is moved toward the soma, the branch accumulates junctions separating sub-branches, and object-identity information is replaced by coincidence information.  At each such sub-branch junction, the post-junction sector of the dendrite is faced with the task of reducing VFE by distinguishing ``true'' from ``random'' coincidences.  Learning and remodeling driven by trophic reward remain the tools available for amplifying the former at the expense of the latter, thus increasing the signal-to-noise ratio and rendering signal processing effectively more digital.  Hence one can expect neurons to remodel in the direction of segregating inputs to their dendritic trees in a way that maximizes true coincidences at branch junctions.  To the extent that they do this, they increasingly operate as hierarchies of spatially-segregated QRFs

Dendritic geometry gives proximal synapses more influence over somatic activity than distal ones; the gating function of proximal inhibitory synapses, in particular, is well known \cite{spruston:08}.  Distal signals can only be processed with high (Bayesian) precision in periods of proximal silence.  Hence it is natural to think of dendritic geometry as imposing a salience gradient from proximal to distal synapses.  Stereotypical differences in arborization pattern between pyramidal cells from different cortical or hippocampal layers reflect different salience assignments, with cortical cells typically giving distal inputs, and hence distal presynaptic partners, higher salience than do hippocampal cells \cite{spruston:08}; see \cite{deitcher:07, mohan:15, beaulieu:18} for specific differences between cortical cells from different layers.  Higher distal salience is only useful from an information processing perspective if it is low-noise, i.e. if PSPs from distal dendritic branches, e.g. from the elaborate apical tufts of Layer V cortical pyramidal cells, are (primarily) time-convolutions of true coincidence signals.  As remodeling takes time, one can expect neurons that compute the same function over long periods, i.e. that are required to exhibit only slow learning, to assign the greatest salience to distal inputs.

In the spiking neurons of interest here, somatic signal integration results, given sufficient amplitude, in AP generation.  While APs are relatively high-amplitude, low-noise signals, outside of myelinated axonal segments differential AP degradation can be expected to contribute to differential presynaptic signal strength.  Weak presynaptic signals are weak actions on the signaling neuron's environment, and cannot be expected to yield strong confirmatory signals in return.  Hence relatively weak synapses can be expected to be clustered, consistent with Fig. \ref{incoming-correlates-fig}b.

\subsection{Neurons as tomographic computers}

As discussed in \S\ref{QRFs}, a QRF corresponds to a specific choice of basis for a subset of the operators $M^A_i$ deployed by an agent $A$ to measure the state of its MB, or boundary $\mathscr{B}$.  We can, therefore, think of the hierarchy of QRFs implemented by a neuron's dendritic tree as a hierarchy of basis choices, with the ``low-level'' basis components corresponding to detections of correlated inputs from local clusters of synapses from single presynaptic partners and the ``high-level'' basis components corresponding to coincidences between increasingly more highly ``chunked'' aggregates of input signals.  Each neuron can thus be thought of as deploying a specific hierarchy of basis vectors to measure the presynaptic activity of its environment.

Consider now a set {\bf N} of neurons sampling the ``same'' environment, e.g. a set of Purkinje cells each sampling the axonal outputs of a set of cerebellar granule cells.  The causal antecedents of the activity of the input neurons are clearly ``unknowns'' from the perspective of the neurons in {\bf N}; the input activities can, therefore, be considered from this perspective an ensemble {\bf E} of samples from an unknown, time-varying state.  We can, in this case, think of {\bf N} as making a set of measurements, each in a distinct basis, of the ensemble {\bf E} of time-varying states.

If we think of the states in {\bf E} as quantum states, then the set {\bf N} of neurons can be regarded as performing quantum state tomography on {\bf E}, and hence quantum process tomography on the process generating time variation in {\bf E} \cite{nielsen:00}.  Quantum state tomography generalizes classical tomography by generalizing the choice of basis away from ``slices'' of a three-dimensional geometry in which the system of interest is embedded.  It is expensive in terms of basis dimensionality, requiring $d^2$ basis vectors for quantum states of (binary) dimension $d$, and $\sim d^4$ basis vectors for processes on states of dimension $d$.  This dimensional cost translates well to the case of ensembles of neural activity, since as in the case of quantum states, it is phase correlation information that the tomographic measurement aims to extract.

Viewing neural computation as tomographic provides an immediate explanation for the enormously high input and output bandwidths and apparent redundancy of neuronal architectures, which are difficult to understand on the basis of simple logic-circuit models \cite{harris:15, luo:21}.  An input space with 100-bit states ($d$ = 100) would require $\sim 100^{4} = 10^8$ binary dimensions for process tomography, or roughly $10^5$ neurons at 1,000 distinct presynaptic partners per neuron.   Following \cite{johansson:07} and assuming $\sim 10^8$ minicolumns with 100 neurons each in human neocortex, analyzing a 100-bit state would require a minimum of $\sim 10^3$ minicolumns or $0.001\%$ of neocortical capacity.  We can infer from this that tomographic computation is approximate, with substantial reduction of the input dimension by salience systems, e.g. active attention.

\subsection{Scaling upwards: minicolumns to functional networks}

The stereotypically layered, only sparsely interconnected minicolumns of mammalian cortex, and in particular, human neocortex, are widely viewed as computational as well as neuroanatomical units, with modeling studies increasingly suggesting that minicolumns execute Bayesian predictive coding \cite{bastos:12, hipolito:21, peters:17, shipp:16}.  Extending the model outlined above, we can think of minicolumns as functionally analogous to neurons, gathering input from and sending output to other minicolumns.  As does a neuron, a minicolumn ``sees'' a spatially-distributed collection of (positive or negative) excitations, spatial and temporal correlations between which are informative to the extent that they are non-random.  Hence a minicolumn is faced with a coarse-grained version of the VFE minimization problem faced by neuron: that of disambiguating ``true'' coincidences from random ones.  Minicolumns can be expected to functionally remodel to increase signal-to-noise ratio for the same reasons neurons can, with trophic reward from the environment as the selective criterion.

Functional networks spanning multiple cortical regions, e.g. sensory pathways, impose a higher-level hierarchical organization.  Predictive coding across layers of this higher-level hierarchy have been analyzed in terms of typical connections between minicolumns in adjacent layers; see e.g. \cite[Fig. 4]{peters:17} or \cite[Fig. 2]{shipp:16} for interlayer connection maps.  Top-down connections encoding likelihood at pathway layer $i$ arise mostly from minimcolumn Layer V pyramidal cells and target minicolumn Layer III cells at pathway layer $i-1$.  Reciprocally, bottom-up connections arise mainly in minicolumn Layer II at pathway layer $i$, and project to networks of Layer IV cells at pathway layer $i+1$. Hence at any level of the pathway, empirical priors are localized in minicolumn Layer III, and likelihoods (or predictions) in minicolumn Layer V. This is further analyzed in \cite{hipolito:21} in terms of how MBs influence connectivity of microcircuits. Internal and external states are implemented by spiny stellate cells and interneurons of each minicolumn; via connections from and to these, respectively, superficial pyramidal cells of the next minicolumn become the (blanket) active states, while the deep pyramidal cells of the previous minicolumn become the (blanket) sensory states. Effectively, the minicolumns are the functional units comprising an MB of networks (either seen as e.g. a MB of MBs, or a Matryoshka of MBs). Once the MBs are functional, the overriding principle is that neurons, microcolumns and networks appear to dynamically self-organize thanks to the FEP. There are many interesting outcomes: consider for instance the visual network as composed of internal states influencing the dorsal and ventral attention networks, while the default-mode network plays the role of a sensory system mediating the influence between these former and external, sensorimotor states. Just as in the case of lower-level structures, such high-level, multi-layer processing pathways exhibit activity-dependent trophic reward and remodel as necessary to achieve it, as shown e.g. by studies of large-scale pathway remodeling following injuries that remove expected inputs \cite{chen:02} (cf. \cite{peters:17, demekas:20}).

Since Friston's proposal that the mammalian brain is an active-inference device \cite{friston:10a}, numerous studies have explored the implementation of active inference by large-scale networks, up to and including the global neuronal workspace \cite{fg:20, fg:21}.  What we have shown here is that these principles extend downward to the scale of the individual synapse.  Trophic rewards select for informative activity, i.e. high signal-to-noise ratios, at every scale.  We can expect these considerations to generalize from neurons to non-neural cells, and from neural communication to communications between biological structures at every scale.

\section{Generalizing neural computation to cellular computation} \label{cells}

The prior discussion was framed in the context of neurons; however, it becomes more generally applicable when we note that neurons did not appear de novo but in fact evolved slowly from other cell types which already shared many of their features (reviewed in \cite{fbl:20}). Not only are the basic molecular components of neurons (ion channels, electrical synapses, neurotransmitter machinery) already present in most cells including unicellulars, but all cells produce bioelectric gradients and comprise tissues with propagating changes in resting potential \cite{levin:17}. It has been suggested that developmental bioelectricity is the evolutionary precursor to neural dynamics, and indeed that evolution speed-optimized the slow bioelectrical signaling that was first used to solve problems in morphospace (exerting anatomical control over body shape by regulating cell behaviors) before it was exapted to solve problems in 3-dimensional behavior space (by regulating muscle function) \cite{levin:19, levin:14}.

Consistent with this hypothesis, bioelectric signaling in non-neural cells has been implicated in control of morphogenetic decisions in embryogenesis, regeneration, and cancer \cite{levin:21b, bates:15, harris:21}. Recent work targeting the native ion channels and gap junctions in tissue has shown that bioelectric states can be readily modulated to predictably alter organ identity, induce regeneration of limbs, control the axial patterning of whole bodies, and repair complex structures such as craniofacial birth defects \cite{mathews:18}. Indeed, it has been suggested that parallel to the architecture of brains, non-neural bioelectricity is the medium implementing the information processing that enables collective intelligence of cellular swarms during body construction and remodeling \cite{levin:21c}.

Morphological change is a deeply computational process that relies on calibrated measurement, and hence QRFs, at multiple organizational levels. While the DNA determines the micro-level hardware that each cell gets to have (e.g., its complement of ion channels), genomes do not directly encode morphology. Instead, they specify machines that have to deploy considerable intelligence, using William James' definition of intelligence as the ability to reach the same outcome from different starting conditions and despite perturbations. The physiology of cellular collectives implements very robust problem-solving capacities by making important decisions about collective macrostates that are not defined at the level of individual cells. For example, mammalian embryos cut in half do not result in two half-embryos – the system regulates to make complete, normal bodies of monozygotic twins (regulative development). Tadpoles that are developmentally disrupted to have all of their craniofacial organs in the wrong positions still result in normal frogs after metamorphosis because the eyes, nostrils, mouth, etc. move through novel paths to get to the same invariant outcome -- a correct frog face \cite{vandenberg:12, pinet:19a, pinet:19b}. Salamanders whose arms are amputated at the shoulder or wrist regenerate precisely what is needed and then stop when a correct limb is complete. These are just a few examples of a nearly ubiquitous property of morphogenetic systems: anatomical homeostasis toward a specific pattern memory, and the ability to reach that state despite sometimes drastic, unpredictable perturbations. This degree of anatomical control requites fundamentally computational functions by cell collectives: they need to be able to measure the current state (e.g. the length of a limb, configuration of the face, etc.), remember the correct state (represent aspects of the correct target morphology), and execute a kind of means-ends analysis to reduce error by controlling cell proliferation, migration, and differentiation.  Trophic reward plays a critical role in such processes: structures or bodies that are functionally insufficient to obtain nutrients and other resources from their environments do not survive.

Consistent with a functional continuity between neural and non-neural dynamics, morphogenesis shares a number of major features with behavior in addition to the reliance on bioelectric networks to implement large-scale coordination. The first is the hardware/software distinction: genomes do not encode final outcomes, they encode the structure of a system with plasticity, context-sensitivity, and the flexibility to produce different outcomes from the exact same hardware. This is why genetically wild-type flatworm cells can generate head structures appropriate to other species when their bioelectric signaling is shifted to different attractors \cite{emmons:15}, why normal skin cells liberated from frogs spontaneously self-assemble into different, motile proto-organisms (``Xenobots'') without any genomic editing \cite{kriegman:20, blackiston:21}, and why embryos with severe genetic defects can be bioelectrically coaxed to normal brain morphogenesis by reinforcing specific voltage patterns \cite{pai:18, pai:15}.  Developmental bioelectric circuits also feature a kind of re-writable memory. In addition to the default voltage patterns (like the ``electric face'' \cite{vandenberg:11, pezzulo:21}), new ones can be written into tissues and maintained by the circuit, such as the two-headed planaria that result in genetically wild-type worms when a different $V_{mem}$ pattern memory is temporarily incepted into the tissue \cite{durant:17, oviedo:10}. These two-headed worms continue to regenerate as two-headed in perpetuity (without additional treatment) or can be bioelectrically switched back to one-headed \cite{durant:17, durant:19}, illustrating the stable but re-writable aspects of the information in the tissue that guides behavior.

Bioelectric signaling in all tissues, like in nervous systems, integrates information across distance to enable decisions and behavior toward adaptive outcomes in novel circumstances. This basic scheme was already discovered by evolution as far back as the time of bacterial biofilms \cite{koshland:83, prindle:15}, and is exploited very widely across the web of life from microbes to humans \cite{levin:21a}. Many aspects of cognitive neuroscience have clear parallels in developmental biology \cite{pezzulo:15}, suggesting that much of the reasoning in \S\ref{macromolecular}--\ref{neurons} above applies not just to nervous systems but in fact to all cells solving problems in various spaces. Indeed, all living systems are deeply hierarchical, exhibiting aspects of basal cognition and decision-making at levels including molecular pathways \cite{watson:10, biswas:21}, physiological problem-solving by cells \cite{emmons:19, jacob:04}, and tissue and organ memory \cite{goel:13, zoghi:04, chakra:97} among others.

\section{Conclusions and future directions} \label{conclusion}

Neurons, networks of neurons, and biological structures generally exchange information with their environments via physical interactions interpretable in terms of measurement (or the gathering of sensory input) and manipulative action.  This interpretation of biological activity forms the basis of the active-inference principle \cite{friston:10a, friston:13a} and has achieved wide currency in the biophysics, evo-devo, and neuroscience communities.  We have shown here how to formulate this view of biological activity in the very general yet powerful language of quantum information theory, employing in particular the idea of a QRF as a calibrated measurement, and in the output direction, a calibrated action.  These results extend the analysis of QRFs as hierarchical Bayesian systems developed in \cite{fg:19b, fg:20, fg:21, fg:20a}.  They allow us to model processes implemented by neurons and networks of neurons, from the biomolecular pathway scale upward, as hierarchies of QRFs.  This representation makes explicit where and how classical information is encoded on successive layers of MBs, enabling models that explicitly comply with energy-budget considerations.  Such models require that cells employ quantum coherence as a resource for bulk computation \cite{fl:21}.  Hence the current framework represents neurons as explicitly quantum devices.

Our goal here has been to develop a framework for building detailed models of specific neuronal cell types in specific environments; these will be pursued in future work.  Even at the current abstract level, however, we are able to predict generically that dendritic trees will remodel in the direction of maximal informativeness, with trophic reward as the selection criterion.  While this prediction is generally supported by the phenomenology of dendritic remodeling \cite{hogan:20, runge:20}, it is specifically testable by correlating measures of branch level activity and branch level active transport.  As branch level remodeling is now known to be involved in neurodegenerative conditions \cite{hogan:20, runge:20}, an understanding of the relations between architecture, computation, and trophic reward at this scale may be useful in ameliorating these conditions.

From a more general perspective, treating neurons in explicitly quantum-theoretic terms introduces new possibilities for mathematical modeling, e.g. with topological field theory as discussed in \S\ref{macromolecular}.  Neuroscience and computer science have both benefitted from the abstraction of neurons to sum-threshold units connected in layered ANNs.  As interest in and tractable architectures for quantum computing continue to develop, we may expect a comparable, but considerably deeper, connection between biological and artificial computing systems.

\section*{Acknowledgements}
ML gratefully acknowledges funding from the Guy Foundation and the Finding Genius Foundation.

\section*{Conflict of interest}
The authors declare no competing, financial, or commercial interests in this research.

%%%%%%%%%%%%%%%%%%%%%%%%%%%%%%%%%%%%%%%%%%
\end{document}